# Forecasting AI Progress: A Research Agenda

Ross Gruetzemacher*, Florian Dorner[†], Niko Bernaola-Alvarez[‡], Charlie Giattino[§], David Manheim[**]

*Auburn University, [†]Freie Universität Berlin, [‡]Universidad Politécnica de Madrid, [§]University of Oxford, [**]University of Haifa

**ABSTRACT:** Forecasting AI progress is essential to reducing uncertainty in order to appropriately plan for research efforts on AI safety and AI governance. While this is generally considered to be an important topic, little work has been conducted on it and there is no published document that gives an objective overview of the field. Moreover, the field is very diverse and there is no published consensus regarding its direction.

This paper describes the development of a research agenda for forecasting AI progress which utilized the Delphi technique to elicit and aggregate experts' opinions on what questions and methods to prioritize. The results of the Delphi are presented; the remainder of the paper follows the structure of these results, briefly reviewing relevant literature and suggesting future work for each topic. Experts indicated that a wide variety of methods should be considered for forecasting AI progress. Moreover, experts identified salient questions that were both general and completely unique to the problem of forecasting AI progress. Some of the highest priority topics include the validation of (partially unresolved) forecasts, how to make forecasts action-guiding and the quality of different performance metrics. While statistical methods seem more promising, there is also recognition that supplementing judgmental techniques can be quite beneficial.

# 1. Introduction

Sufficiently advanced AI has the potential to radically transform society in the coming decades. This societal transformation from AI could be either very good for humanity, very bad for humanity or some strange mix of good and bad. For example, these technologies could dramatically increase life expectancy, slow aging, rid the world of many mortal illnesses and increase economic productivity greatly increasing wealth for a large portion of the population. Alternately, superintelligent AI could lead to existential catastrophe (Bostrom 2014). More likely than these extreme cases are futures that involve both positive and negative outcomes: e.g. life expectancy increases greatly while aging slows but income inequality grows as the consequence of extreme labor-displacing AI. However, no future is set in stone, and policy makers now have the opportunity to shape the future for the billions.

In order to mitigate risks from AI and to maximize the potential for positive futures, there is tremendous value in reducing uncertainty regarding timelines of AI progress. Doing so can enable decision makers in governments and major organizations to make better decisions with respect to these issues, and it can help researchers working on issues of AI safety and AI governance to prioritize their own research goals. For this latter reason, the topic of AI forecasting is of great interest to these researchers, who are among the best placed to understand the uncertainties and critical gaps in knowledge, yet there have been no objective efforts to clarify the major topics of interest. The current paper addresses this, and uses expert understanding to inform a research agenda for advancing the body of existing literature on the topic.

Technological forecasting in general is a challenging research area, and forecasting AI progress specifically is even more challenging. Some of the particular challenges posed by AI forecasting include the difficulty of measuring progress and the breadth and ever changing nature of the types of different applications of AI (e.g. self-driving cars, the generation of synthetic media, personalized medicine). More fundamentally, the nature of "intelligence" itself is difficult enough to define and measure in humans, let alone in machines. Unfortunately, the methods used in psychometrics do not apply to assessing progress in AI, though significant research has gone into trying to develop a similar framework which would apply to both human and machine intelligence (Hernandez-Orallo 2017). However, another way in which forecasting AI progress is unique lies in the fact that while the focus is intelligence, most forecasts are concerned with one of the seemingly myriad subdomains rather than the broad objective of the field. To be certain, this may be true for other technologies, but for AI, many of the subdisciplines have the potential transformative impact of general purpose technologies (GPTs; Bresnahan and Trajtenberg 1995) independently, and these subdisciplines can be further segmented into valuable forecasting targets. One of the most highly cited papers related to this concerned the forecasting of the automatability of nearly 1,000 different human occupations[1] (Frey and Osborne 2013). Technologies like biotechnology and nanotechnology can be thought of as similar to AI in that they are also GPTs, but neither of these examples have such a broad potential social and economic impact. It has been suggested that AI broadly is a general purpose principle (Lipsey et al 2005) rather than a specific technology, and that specific AI technologies (even those considered to be "narrow") can independently constitute GPTs. Moreover, abstract notions of artificial general intelligence (Goertzel 2007) have the potential for even more radical transformative impacts (Gruetzemacher and Whittlestone 2019).

---

[1] This is an example of *future of work* research which is mentioned in section 3.3.

These issues make it more challenging to model AI progress in the same way as other technologies. This could also perhaps explain the dearth of broader AI forecasting studies published in the leading academic outlet for technology forecasting[2].

Despite these challenges, there are reasons for optimism that a concerted, holistic research effort could help reduce our uncertainty about future AI progress. For instance, some relevant trends have been impressively regular and quantifiable over the long term, such as computer hardware (e.g. Moore's Law[3]) and performance in specific domains (e.g. chess[4]), and while some aspects of AI progress are more difficult to quantify, recent forecasting work aggregating human judgment in the geopolitical domain suggests that even less-quantifiable aspects might be amenable to accurate forecasting (Tetlock and Gardner 2016). Any research agenda attempting to forecast AI progress will need to integrate insights from these different approaches to be most effective.

However, Russell and Norvig's forecast (see footnote 4 below) is an exception, and most previous attempts to forecast AI progress have not been very effective (Armstrong et al. 2015). Perhaps this is because most previous forecasts have attempted to forecast targets that were much less specific than the performance on a widely accepted metric for measuring human performance in chess. The most common target of previous forecasts has been the general notion of machine intelligence (Michie 1973) or so-called high level machine intelligence (HLMI; Muller and Bostrom 2016; Grace et al. 2018). There are no metrics for measuring progress toward such ambitious forecasting targets, although there is substantial and ongoing work in this area (Hernandez-Orallo 2017). Moreover, other forms of advanced AI have the potential to dramatically or radically transform society (Drexler 2019). Little effort has been made to forecast progress toward such futures, or their relative likelihood.

To help address the challenges of AI forecasting and to provide a starting point for this nascent field, we present a research agenda for forecasting AI progress. To our knowledge, this is the first such attempt. To help set this agenda, we conducted a Delphi study in which a diverse group of experts in the field identified and ranked important research questions and methods. This study will proceed by very briefly reviewing forecasting and AI forecasting literature, then discussing the Delphi process used for generating the research agenda. We then report the results, which are broken up into important questions and suitable methods. *Each subsection for the questions and methods includes a subsubsection which identifies concrete research suggestions for future work.* These sections are followed by a discussion of the results and limitations, and finally, a section highlighting the conclusions of the study.

---

[2] Technological Forecasting and Social Change.
[3] Gordon Moore, while working at Fairchild Semiconductors in the 1960s, famously tracked numerous different parameters: cost per transistor, number of pins, logic speed. After several years, it became clear that the number of transistors per chip was fitting a nice curve. The success of this curve, which would come to be known as Moore's Law, at predicting the progress of the semiconductor industry led to its official adoption by the Semiconductor Industry Association for inclusion in a formal technology roadmap for the industry.
[4] Russell and Norvig in the first edition of their classic AI textbook (Russel and Norvig 1995). Here, they plotted the ELO score of the best chess performing algorithms starting in 1965 and extrapolated, predicting correctly that an algorithm would surpass expert human level (i.e. Gary Kasparov) in 1997. Moreover, games have long been used as a means of measuring progress in AI (Samuel 1959), although separating the signal from the noise of these indicators is often challenging.

## 2. Background

A full literature review of relevant forecasting methods is beyond the scope of this paper. However, such a literature review would complement this research agenda well and, if well done, would be a valuable contribution to the study of forecasting AI progress. Thus, we recommend this for future work. Here we simply attempt to conduct a very brief review of the literature merely to add some context for readers who may not be familiar with either forecasting in general or previous AI forecasting efforts.

There are broadly two primary classes of forecasting techniques: statistical and judgmental (Armstrong 2001). Statistical techniques are prefered for more applications (Hyndman and Athanasopoulos 2018), however, judgmental techniques are still more appropriate for a variety of different purposes (Tetlock and Gardner 2016; Green et al. 2015). There are also auxiliary techniques, such as scenario analysis, which do not fall into either of these categories (Roper et al. 2011). Such techniques are more common for technology forecasting. Tech mining (e.g. bibliometrics and scientometrics) is a particular form of statistical forecasting techniques which is widely used for applications in technology forecasting (Daim et al. 2016; Porter and Cunningham 2005).

The most common and widely used statistical forecasting technique is widely thought to be trend extrapolation (Roper et al. 2011). While simple, the technique is also very powerful and can be very effective (e.g. the example previously discussed from Russell and Norvig, 1995). Indicators are variables for which data exists that can be used to extrapolate trends which have implications for future progress in some dimension relevant to the thing being forecast. Other common statistical forecasting techniques include econometric modeling and machine learning based techniques (Hyndman and Athanasopoulos 2018).

Judgmental forecasting techniques are commonly employed for a variety of different tasks related to AI forecasting. The most common forms of expert elicitation for forecasting are interviews and surveys. Some other common techniques include the Delphi (Helmer 1960), prediction markets (Wolfers and Zitzewitz 2004) and forecasting tournaments (Tetlock and Gardener 2016). Targets are the thing which is being forecast, and commonly need to be well specified and unambiguously evaluable in order to be valuable.

Forecasting AI progress has been a topic of interest since the inception of AI at the 1958 Dartmouth Conference where attendees were polled about future progress (Muehlhauser 2016). Another well known early example of attempts to forecast advanced AI is that of Michie (1973) who conducted a survey following a lecture. Since 2006 there have been twelve major surveys among experts and non-experts (Zhang and Dafoe 2019; Grace 2015). Five of these have been academic studies involving experts (Baum et al. 2011; Muller and Bostrom 2016; Grace et al. 2018; Walsh 2018; Gruetzemacher et al. forthcoming).

Surveys may make up a large amount of the existing literature concerning AI forecasting, but they certainly do not account for all of it. There are a number of analyses that have been conducted assessing the viability of forecasting AI progress (Armstrong et al. 2015) as well as previous unpublished efforts (Muehlhauser 2016), and a large, growing and varied body of work on the topic has been conducted by the nonprofit organization AI Impacts[5]. Significant work has also been done to identify measures of machine intelligence (Hernandez-Orallo 2017) as well as for identifying methods for modeling AI progress (Brundage 2016) and indicators of AI progress beyond performance measures (Martinez-Plumed et al. 2018). Aside from surveys, perhaps the most significant forecasts have been trend extrapolations of different indicators such as computational resources required for training groundbreaking AI models (Amodei and Hernandez 2018), investment into large AI research projects (Gruetzemacher 2019b) or computational efficiency of algorithms for replicating the results of past milestones (Hernandez and Brown 2020). The most noteworthy example of this was mentioned earlier when Russell and Norvig (1995) used the technique to predict superhuman performance in chess.

---
[5] www.aiimpacts.org - the scope of the work conducted by this organization is too broad to discuss in detail here, however, the most practical forecast that has been generated was the Grace et al. (2018) survey.

Other efforts that are related to AI forecasting are *future of work* studies which involve forecasting future labor markets which are assumed to be significantly impacted by automation from AI. A seminal study on the topic was conducted by Frey and Osborne (2017) which used a novel technique to project that 47% of jobs were susceptible to automation in the coming decades. Surveys (Duckworth et al. 2019) as well as data based methods (Das et al. 2020; Martinez-Plumed et al. 2020) have also been used for such forecasts. There is also a significant body of work among machine learning researchers to develop benchmarks for rapid progress in different research domains (e.g. natural language processing; Wang et al. 2018; Wang et al. 2019; Zellers et al. 2020).

## 3. Delphi Process

For developing this research agenda we utilize the Delphi technique to elicit and aggregate experts' opinions regarding the best questions and methods to prioritize when forecasting AI progress. The Delphi technique is most commonly used for forecasting directly (Rowe and Wright 2001), but can also be used in other ways such as the policy Delphi in which it is used to generate opposing views of a topic (Turoff 1970). Despite originally developed at the RAND Corporation for forecasting (Helmer 1967), the Delphi technique is a general tool that has been used previously for generating research agendas in a variety of disciplines (Kellum et al. 2008; Dahmen et al. 2013) including medicine (Burt et al. 2009), art therapy (Kaiser and Deaver 2013) and school counseling (Dimmit et al. 2005). We developed a customized Delphi process to meet the specific objectives of this research agenda, summarized below. A more complete description of the methodology can be found in Appendix A.

Based on the previous studies which used the Delphi technique for eliciting and aggregating expert opinion of salient research topics, we chose to first use the Delphi technique for identifying experts' opinions of the most important research topics and then for rating the importance and feasibility of these topics. Previous studies have used the Delphi similarly: One had an additional initial round where research goals were identified (Dimmit et al 2005) and in another one, research topics were ranked over two rounds after topics had been identified (Gordon & Barry 2006). The Delphi process that we used is illustrated in Figure 1. It begins with the distribution of a Delphi questionnaire, consisting of four questions (see Table 1), where responses had no length requirements or limits. We next summarized and aggregated the questionnaire responses by, for example, deduplicating equivalent responses and linking together common themes. Then we reported the summarized responses back to the first-round participants for comments and discussion. Following this, we distributed the questions and methods to the participants who had completed the first round for scoring.

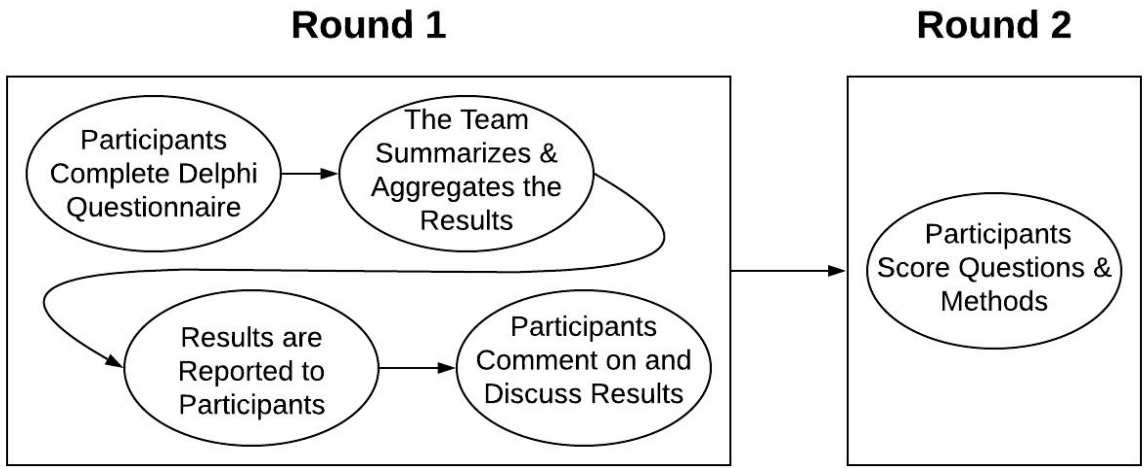

**Figure 1:** An illustration of the Delphi process used here.

Of critical importance when conducting any Delphi study is the selection of experts. In this study we primarily considered experts who had previous experience in either economics, technological forecasting or AI forecasting. We also invited two experts who had substantial experience in the use of foresight techniques for AI (i.e., workshops or scenario planning techniques). These experts were representative of academia, government, industry and nonprofits. 15 experts of 32 responded to our invitations to participate, a rate similar to previous studies (Beddoe et al. 2016).

Aside from the experts, the other most significant component of a Delphi study is the content of the Delphi questionnaire. The studies considered here which have previously used the Delphi for creating research agendas are variants which focus on planning, and we drew inspiration from literature concerning this application in our design process (Linstone and Turoff 1975). For this study, four questions were included in the questionnaire. These four questions can be seen in Table I below:

## Table I: Delphi Questionnaire

| | |
|---|---|
| 1. | Do you feel that forecasting AI progress is, or could be, a well defined research topic? Why? |
| 2. | What questions should researchers who work on forecasting AI progress prioritize? |
| 3. | What methods or techniques should researchers use/prioritize to answer these questions? |
| 4. | Are there any topics relevant to forecasting AI progress that you feel are important but neglected? Why? |

Following the conclusion of the first round, responses to the first question were summarized and lists of questions and methods were compiled and provided to participants. In the second round, participants then rated the questions and methods on a scale from 1 to 5 along the axes of importance and feasibility. 12 out of 15 of the participants completed this section, which is in line with previous studies (Gordon and Barry 2006).

The Delphi process was led by four facilitators. The lead facilitator was in charge of inviting and contacting the participants through email. The remaining three facilitators each participated in the design of the custom Delphi process utilized as well as in developing the questionnaire and the mechanisms for facilitating the process. All facilitators also contributed to the summarization and aggregation of the results from the first round and the lead facilitator and two of the co-facilitators participated in the analysis.

The questions are reported in the results as they were reported to participants after the 1st round of the Delphi: they are structured in clusters descending from the topics which were perceived by the facilitators to have been of the most interest to participants[67]. Each cluster involves a question which was perceived as a more general question that encompassed to some degree each of the sub-questions comprising the cluster. Each cluster was labeled for the purpose of presenting the results, but these labels were not reported to the participants of the Delphi with the outline and other results from the first round.

# 4.   Delphi Results

The results of the second round of the Delphi are presented in Tables II, III and IV. For the sake of brevity and to focus on the research agenda, we do not discuss these results in detail[8]. The research agenda in the following section is based on the results of the Delphi and is reported following the outline of the questions and methods that are reported in Tables II and IV. Table III simply identifies the names of each of the clusters identified in Table II.

Table II shows the research questions of interest organized by group and cluster, as previously described. The questions that are marked in italics are the primary question for each cluster. The final cluster were questions which did not fall into either of the other two groups or into any of the other nine clusters. These are marked as miscellaneous. The three groups are marked on the left of Table II: meta-level topics, forecasting methods-related topics, and dissemination and miscellaneous topics. The mean importance and mean feasibility columns are of primary interest, and are shaded in grey. In each of these columns, all italicized numbers indicate that these results are greater than the mean: 3.83 and 3.35 for the importance and feasibility, respectively. All bold numbers indicate that these values fell in the top 10 results for the column.

Table IV depicts the methods of interest organized by group. The columns are all the same as in Table II except for the lack of a column for clusters - no clusters were identified for the methods. The text formatting in this table indicates the same relationships within the data as it does for Table II (i.e. bold numbers are in the top 10 and italicized numbers above the mean). The mean for importance and feasibility for the methods scores were 3.64 and 3.88, respectively.

Regarding the clustering of questions by topic: we openly acknowledge that some questions may be good fits for more than one cluster or more than one group, but we feel that no questions are grouped or clustered in a manner where the authors' reasoning can not be inferred. Moreover, we stress that the following sections represent experts' scores on questions that one or more had identified to be the most important. Thus, we suggest that poor scores on these questions do not indicate that the question is unimportant or unsuitable for future work. We hope that all questions presented here will be perceived as important topics to explore, and we encourage any reader who agrees or who is otherwise inspired to pursue one of these questions to do so.

---

[6] More detail regarding the Delphi procedure and its implementation can be found in Appendix A.
[7] Based on frequency; the complete outline described here is presented in Appendix B.2.
[8] Further discussion of these results is included in Appendices C and D.

## Table II: Questions Scored by Mean Importance & Feasibility

| Group | Cluster # | Question | Mean Importance | Median Importance | SD Importance (# responses) | Mean Feasibility | Median Feasibility | SD Feasibility (# responses) |
|---|---|---|---|---|---|---|---|---|
| Meta-level Topics | 1 | *What are the most important forecasting targets?* | | | | | | |
| | | How do we define qualitative and quantitative measures of progress toward forecasting targets? | 4 | 4.5 | 1.15 (10) | 3.65 | 3.25 | 0.82 (10) |
| | | How can we decompose abstract AI technologies into more easily forecastable targets? | **4.11** | 4 | 0.7 (11) | 3.64 | 4 | 0.67 (11) |
| | | What questions/targets matter for practical, near-term decision making? | **4.48** | 5 | 0.71 (9) | **3.83** | 4 | 0.61 (9) |
| | 2 | *What are the implications of timelines?* | 3.68 | 4 | 0.95 (8) | 3.14 | 3 | 0.9 (7) |
| | | Should we focus on capabilities or the impact of AI systems? | 3.42 | 3.8 | 1.5 (9) | 2.69 | 2 | 1.33 (8) |
| | | How can forecasts be applied to identifying and mitigating risks? | 4.04 | 4 | 1.01 (9) | 2.88 | 3 | 0.99 (8) |
| | 3 | *How do we best evaluate overall AI progress?* | 3.93 | 4 | 1.14 (11) | 3 | 3 | 1.0 (11) |
| Forecasting Methods-related Topics | 4 | *What are the most useful indicators (e.g. compute, talent, investment/resources, economic impact, benchmark performance)?* | 4.1 | 4 | 0.84 (8) | **3.89** | 4 | 0.64 (8) |
| | | What performance metrics are relevant and most effective? | **4.26** | 4 | 0.77 (7) | 3.54 | 3.8 | 0.96 (7) |
| | | How do we assess the quality of a metric/benchmark's signal? | **4.2** | 4 | 0.68 (9) | 3.44 | 4 | 1.01 (9) |
| | | Are existing (SOTA) benchmarks relevant or useful (i.e. strong signal)? | 3.87 | 4 | 0.33 (9) | **3.78** | 4 | 1.3 (9) |
| | | Should we focus on tasks or abilities for measuring and forecasting AI progress? | 3.22 | 3.4 | 1.48 (8) | 2.38 | 2 | 0.92 (8) |
| | | How would we develop a broader discipline for measuring and assessing progress in AI (like psychometrics)? | 3 | 3 | 1.31 (8) | 3.09 | 3.35 | 1.1 (8) |
| | | How do we best analyze/measure AI systems' abilities to generalize, understand language and perform common sense reasoning? | 3.77 | 4 | 0.83 (9) | 2.77 | 3 | 0.82 (9) |
| | 5 | *How can we model AI progress?* | 3.76 | 3.8 | 0.83 (9) | 3.21 | 3 | 0.82 (9) |
| | | What are the best methods for modeling given the correct variables? | 3.29 | 3 | 1.1 (9) | 3.33 | 4 | 1.22 (9) |
| | | Why is progress faster in some metrics than others? | 3.4 | 3.6 | 1.32 (9) | 3.56 | 4 | 1.42 (9) |
| | | Can independent variables be used to model AI progress effectively model progress in other fields/research domains? | 3.31 | 3 | 0.96 (8) | 3.12 | 3 | 0.83 (8) |
| | 6 | *What are the most probable AI development scenarios?* | **4.11** | 4 | 0.78 (9) | 2.86 | 3 | 0.89 (9) |
| | | How do we identify the most plausible paths for a variety of transformative AI technologies/systems? | **4.2** | 4 | 0.84 (9) | 2.83 | 3 | 0.87 (9) |
| | | What will be the new applications/services made possible by new AI technologies? | 3.41 | 3 | 1.12 (9) | 2.49 | 2 | 0.92 (9) |
| | | What impact does NLP have on AI capabilities? | 3.34 | 3.6 | 1.3 (9) | 3.21 | 3.4 | 1.17 (9) |
| | 7 | *How do we produce the best forecasts?* | 4.03 | 4.3 | 1.23 (9) | 3.06 | 3 | 1.13 (9) |
| | | How do we aggregate and report metrics? | 4.09 | 4 | 0.85 (8) | **4.31** | 4.25 | 0.7 (8) |
| | | What are/how do we develop the best qualitative/quantitative a priori models? | 3.93 | 4 | 1.02 (9) | 2.78 | 3 | 0.94 (9) |
| | 8 | *How effective can long term forecasting of AI progress be?* | 4.1 | 4 | 0.93 (9) | 2.13 | 2 | 0.95 (9) |
| | | How do we best validate forecasts of AI progress: historical data/near-term progress? | **4.62** | 5 | 0.74 (8) | **3.84** | 4 | 0.78 (8) |
| Dissemination and Miscellaneous Topics | 9 | *How do we utilize forecasts to inform decision makers and develop appropriate and measured initiatives/interventions?* | 4.54 | 5 | 0.73 (8) | 3.34 | 3.4 | 0.75 (7) |
| | | Who are the relevant stakeholders/audiences for forecasts and how do we best report forecasts to each? | 3.78 | 4 | 1.3 (9) | **4.24** | 4 | 0.71 (8) |
| | | What are information hazards related to AI forecasts and how do we best make decisions about how to guard and disseminate forecasting data? | **4.17** | 4 | 0.65 (10) | **3.99** | 4 | 1.07 (8) |
| | | What can we learn from historical examples of policy making? | 3.51 | 3.6 | 1.12 (9) | 3.68 | 3.7 | 0.71 (8) |
| | 10 | How can we improve/make more useful conventions regarding forecasting questions and answers? | 3.3 | 3 | 0.67 (10) | 3.51 | 4 | 0.72 (10) |
| | | How do we forecast the automatability of different types of unique human tasks? | **4.12** | 4 | 0.83 (8) | 3.36 | 3.45 | 0.73 (8) |
| | | How can we collect data measuring human performance that can easily be compared to machine performance (e.g. next word prediction log loss)? | 3.52 | 3 | 1.0 (9) | **4.15** | 4.1 | 0.99 (8) |
| | | Can we identify a minimum viable timeline (e.g. 10% of strong AI) for use by stakeholders and decision makers? | 3 | 2.5 | 1.51 (8) | 2.5 | 2 | 1.41 (8) |
| | | What can we learn from existing long-range forecasting techniques (e.g. clionomics, K-wave theory, S-curves)? | 3.71 | 4 | 1.38 (7) | **4.29** | 4 | 0.76 (7) |
| | | How do we best operationalise group forecasting efforts? | 3.69 | 3.75 | 0.7 (8) | 3.62 | 4 | 0.92 (8) |
| | | How effective are existing methods at forecasting technology (e.g. prediction markets, the Delphi)? | **4.14** | 4.2 | 1.0 (10) | **4.02** | 4.1 | 1.06 (10) |

## Table III: Cluster Topics

| # | Cluster Topic |
|---|---|
| 1 | Forecasting Targets |
| 2 | AI Timelines |
| 3 | Evaluating Progress |
| 4 | Indicators and Metrics |
| 5 | Modeling AI Progress |
| 6 | Concrete Scenarios |
| 7 | Improving Forecasting Efforts |
| 8 | Long Term Forecasting |
| 9 | Dissemination |
| 10 | Miscellaneous |

## Table IV: Methods Scored by Mean Importance & Feasibility

| Group | Method | Mean Importance | Median Importance | SD Importance (# responses) | Mean Feasibility | Median Feasibility | SD Feasibility (# responses) |
|---|---|---|---|---|---|---|---|
| Statistical Methods | Statistical forecasting techniques | 3.5 | 3.5 | 2.12 (2) | 3 | 3 | 2.83 (2) |
| | Statistical modeling | **3.9** | 4 | 0.89 (5) | 3.9 | 4 | 0.89 (5) |
| | Extrapolation | **4.07** | 4 | 0.65 (6) | **4.4** | 4.7 | 0.8 (6) |
| | Bayesian methods | **3.88** | 4 | 1.0 (6) | **4.05** | 4 | 0.64 (6) |
| | Benchmarks & metrics | 3.82 | 4 | 0.46 (5) | **4.08** | 4 | 0.73 (5) |
| | Aggregating into metrics for human comparison | **4.2** | 4 | 0.84 (5) | 3.6 | 4 | 0.55 (5) |
| | Item response theory | 3 | 3 | 0.0 (2) | **4.5** | 4.5 | 0.71 (2) |
| | Data science (e.g. tech mining, bibliometrics, scientometrics) | 3.67 | 4 | 1.51 (6) | **4** | 4 | 0.71 (5) |
| | Theoretical models | **3.98** | 3.95 | 0.9 (6) | 3.3 | 3.4 | 1.35 (6) |
| | Machine learning modeling | 3.83 | 4 | 1.34 (7) | 3.67 | 4 | 1.25 (7) |
| | Simulation | 3.59 | 3.85 | 0.9 (8) | 3.5 | 3 | 0.76 (8) |
| Judgmental Methods | Judgmental forecasting techniques | 3 | 3 | 0.0 (2) | 3.5 | 3.5 | 2.12 (2) |
| | Simulation & role-play games | 2.91 | 3 | 1.06 (9) | **4.38** | 4 | 0.52 (8) |
| | Scenario analysis | 3.59 | 4 | 1.12 (9) | **4.08** | 4 | 0.85 (8) |
| | Blue-team/red-team | 3.8 | 4 | 1.3 (5) | **4.4** | 5 | 0.89 (5) |
| | Expert elicitation | 3.38 | 3.9 | 0.88 (5) | 3.75 | 4 | 0.5 (4) |
| | Delphi | 3.69 | 4 | 0.94 (7) | **4.29** | 4 | 0.49 (7) |
| | Expert adjustment | 3.8 | 4 | 0.84 (5) | **4.6** | 5 | 0.55 (5) |
| | Prediction markets | 3.18 | 3 | 1.59 (8) | 3.71 | 4 | 1.11 (7) |
| | Forecasting tournaments | 2.99 | 2.5 | 1.18 (8) | 3.93 | 4 | 0.93 (7) |
| | Calibration training | 2.79 | 3 | 1.15 (7) | 3.98 | 3.95 | 1.08 (6) |
| | Aggregation of expert opinion | 3.34 | 3 | 0.94 (7) | **4.24** | 5 | 1.23 (7) |
| | Immersive observation of AI labs | **3.84** | 4 | 0.99 (8) | 3.03 | 3 | 0.58 (7) |
| | Identifying clear and effective forecasting targets | **4.29** | 4 | 0.49 (7) | 3.67 | 4 | 0.52 (6) |
| | Conceptual progress acceleration survey (using pairwise comparisons) | 3.78 | 4 | 0.44 (5) | 3.7 | 3.9 | 0.48 (4) |
| Other | Hybrid methods (i.e. statistical and judgmental) | **4.52** | 5 | 0.83 (6) | 3.78 | 3.9 | 0.83 (5) |
| | Probabilistic reasoning (e.g. the Doomsday argument) | 3.25 | 3.5 | 1.33 (6) | 3 | 3 | 0.89 (6) |
| | In-depth analysis of specific questions | **4.19** | 5 | 1.19 (8) | **4.17** | 4 | 0.86 (7) |
| | Literature review | **3.88** | 3.5 | 0.99 (8) | **4.43** | 5 | 0.98 (7) |

A discussion of the results of the Delphi process reported here, while not included in the main text of this paper, are still significant. This discussion is included in Appendices C and D. The following section that discusses the research agenda builds on these results. This section is included to act as an accessible reference for readers as they read through the research agenda.

# 5. Concrete Research Directions

Based on the Delphi elicitation, we developed a research agenda comprised of concrete research suggestions, which also incorporates our own relevant areas of expertise and our experience working on the topic. The results of the survey do not themselves directly lead to concrete research proposals and areas. For this reason, there is some synthesis of the topics, methods, and meta-level questions into a research agenda, along with identifying and filling gaps that emerge from this synthesis. The agenda is presented in subsections 5.1 and 5.2 containing suggestions for the salient questions and methods-related suggestions, respectively. Each of these sections is divided into subsections based on groups, clusters of questions and different types of forecasting techniques.

## 5.1 Research Ideas for the Salient Questions

The results from the Delphi process yielded a large number of very valuable questions relevant to this research agenda, reviewed in this subsection. Questions that emerged while synthesizing the agenda also deserve attention, and those which do not fit within any of the existing clusters are described in section 5.1.3 with the other miscellaneous topics. The structure otherwise follows the outline of the most important questions as they were reported to the participants following the first round of the Delphi.

Each of the clusters is characterized by a single question. These can be seen in italics in Table II. For this research agenda, these questions were extended to more comprehensively represent the entire range of questions included in the cluster. In the sections below, as we present the research agenda, we begin the discussion of concrete research proposals for each cluster by highlighting this expanded version of the cluster's primary question in italics.

### 5.1.1 Meta-level Topics

Forecasting Targets[9]

> *Q.1 What are the most important forecasting targets and how can they be designed in a manner that is most effective at identifying valuable information and signal regarding the forecasts of interest?*

Well-defined forecasting targets are crucial for evaluating a wide variety of forecasts and different forecasting techniques. It is not only necessary that these targets are well-defined, but also that they are objectively and unambiguously evaluable, near-term probable and indicative of some signal of progress that is useful to decision makers. While these desiderata outlined by Dafoe (2018) are useful guidelines for creating effective forecasting targets, the creation of these targets in practice remains very difficult (Gruetzemacher et al. forthcoming). Work extending the desiderata proposed by Dafoe is certainly welcome. Of particular interest along these lines would be a careful evaluation of different AI forecasting targets (and their resolutions) that have been used on ai.metaculus[10] or in recent AI surveys (Grace et al. 2018). Also interesting would be an analysis of technological forecasting targets and resolutions from prediction markets or from previous studies such as SciCast[11]. In a slightly different vein, feedback from experts about progress in better defining and forecasting AI developments may be useful, especially for feedback when the forecasting targets in question are still far in the future.

---

[9] A forecasting target is the target of the forecast. i.e. the thing which is being predicted.
[10] ai.metaculus.com.

Decomposition of forecasting targets is widely used in the presence of high uncertainty (MacGregor 2001), and it would be useful to demonstrate steps for effectively using this technique in the context of AI. This could involve an experiment to forecast benchmark performance on some measure, like SuperGLUE (Wang et al. 2019), at a given time (e.g. January 1st, 2021) using a model of two input indicators such as largest trained model size (in parameters) and largest cleaned dataset size (in GB). This would be a relatively simple experiment to carry out (perhaps similar to recent work by Kaplan et al. 2020, but for forecasting.)

The Delphi process utilized for the expert elicitation that was used to create this research agenda was effective at identifying top questions and then ranking them. Due to the highly technical nature of current AI development, using this process for identifying useful and near-term probable AI forecasting targets is something which should be explored further. A straightforward study could be conducted utilizing this technique and comparing the results and targets' resolutions with those on ai.metaculus generated using other techniques.

While the Delphi process demonstrated here may be an effective way to leverage expert opinion to create evaluable forecasting targets, such targets would not necessarily be practically useful for decision makers. Thus, a separate direction for future work could examine technological forecasting targets from previous work, and their resolutions, to determine what might have been most useful for improving decision making. It might be useful to survey decision makers (e.g. policy professionals and executives) regarding their preferences on these past forecasting targets. This could also be done for the smaller body of AI forecasting targets, regardless of whether or not they have resolved. Similarly, a survey could be conducted of decision makers or AI policy researchers. These proposed projects could be useful with small sample sizes and may benefit if the surveys are administered interactively with select experts as structured interviews. Even more straightforward would be to simply use various forms of expert elicitation to obtain decision makers' or AI policy researchers' opinions on the most important forecasting relevant questions.

### AI Timelines

> *Q.2   What are the implications of AI timelines and how can they be formulated such that they maximize benefits and minimize harms?*

Developing timelines for the arrival of radically transformative AI has been a common focus for previous work on forecasting AI progress (Grace et al. 2018; Gruetzemacher et al. forthcoming). The most important implications of these timelines is for mitigating catastrophic and existential risks. For this reason, AI timelines are of significant importance. However, it is likely that their value is diminished because they predict abstract, poorly defined notions (e.g. HLMI). Efforts to create AI timelines for numerous different plausible scenarios would be a welcome research direction, although its tractability likely makes it too difficult. Surveys are valuable, but likely more valuable for other forecasting topics than actionable long-term AI timelines (e.g. short-to mid-term forecasts, identifying important impact areas or research domains).

---

[11] See www.SciCast.org. Other exploration of the results from this dataset may also lead to useful information for improving AI forecasting efforts.

Evaluating Progress

> *Q.3  How do we best evaluate overall AI progress?*

While simple and straightforward, this is perhaps the most important and most challenging question related to the forecasting of AI progress. The most significant efforts to this end have included the aggregation of different indicators and data for mining (Brundage and Clark 2017; Eckersley and Nasser 2018; Martinez-Plumed et al. 2020). However, the importance of this issue is quickly generating increased interest in the forecasting and AI research communities, and 2020's Evaluating Progress in AI workshop at the European Conference on AI marked the first concerted effort to address this question.

One pathway to address this challenge would be through the creation of a broader discipline for measuring and evaluating progress in AI. As psychometrics applies to measuring and evaluating intelligence, a new discipline could apply to intelligence without the anthropomorphic limitations of psychometrics. Hernandez-Orallo (2017) has presented some first steps toward this form of assessment.

## 5.1.2 Methods-related Topics

Indicators and Metrics

> *Q.4  What are the most useful indicators (e.g. compute, talent, investment/resources, economic impact, benchmark performance) and how can we evaluate their signal relevant to different topics of interest?*

Indicators are critical for statistical models of AI progress. Substantive work has already been conducted to identify the most valuable indicators of AI progress (Martinez-Plumed et al. 2018; Martinez-Plumed and Hernandez-Orallo. 2018). Recent work has gone further to demonstrate the use of scientometrics to obtain indicators of institutions' relative AI research performance (Barredo et al. 2020). Further work should be conducted to identify salient indicators using scientometrics that can be used in different statistical forecasting models.

Substantial work on indicators already exists in the technological forecasting literature (Porter and Cunningham 2004), and there are likely many possible applications of this existing work in the context of forecasting AI progress. Furthermore, it is likely that research on technological forecasting, particularly that using scientometrics to identify rapid growth or to project accelerating research progress, would be very useful to those working to forecast AI progress. For example, recent work from Klavans et al. (2020) has significant implications for AI forecasters. Any work in this vein is welcome, and efforts to evaluate the technique in the context of forecasting AI progress would likely also make for valuable contributions.

Modeling AI Progress

> *Q.5  How can we best model progress in AI and what can we learn from previous work in other disciplines about problems in forecasting AI progress such as modeling potential discontinuous progress?*

Modeling AI progress is an ambitious goal that has seen few efforts. Brundage discussed the possibility of modeling AI progress (Brundage 2016) and some simple extrapolative models have been proposed (Amodei and Hernandez 2018; Gruetzemacher 2019b). More complex models of multiple inputs, like that proposed by Brundage, are of interest here. Work on this topic is likely challenging, and any progress on the topic is welcome.

One particular challenge in modeling AI progress is modeling discontinuous progress. AI Impacts has conducted extensive work on historical discontinuous progress in technological development[12]. Gruetzemacher (2019a) has proposed adaptations of Monte Carlo simulation to address these issues in hybrid forecasting processes; this technique could also be applied for statistical models of AI progress. Alternately, work to incorporate models of discontinuous technological progress, such as that of Klavans et al. (2020), into more complex models of broader AI progress would be welcome contributions to the community.

### Concrete Scenarios

> *Q.6 What are the most likely scenarios for the development of transformative AI or radically transformative AI, and how can we best foresee potential future capabilities and applications (e.g. natural language processing or robot learning)?*

Mapping the technological landscape is a critical element of the AI governance research agenda (Dafoe 2018), yet little work on this topic is publicly known[13]. Gruetzemacher (2019a) proposed a variety of scenario-based techniques, dubbed scenario mapping techniques, for this purpose. Gruetzemacher more recently has provided a more detailed explanation of these techniques and their application (Gruetzemacher 2020). While this work was extensive, there are numerous novel techniques worth exploring. Interested readers could look to the specific holistic forecasting framework proposed by Gruetzemacher, and attempt to create variations. Alternately, entirely novel methods are also welcome. It is likely possible to generate plausible scenarios by combining powerful bibliometrics and scientometric analyses with expert adjustment of some sort; this is likely a challenging but valuable area of research.

### Improving Forecasting Efforts

> *Q.7 How do we improve the aggregation of data and opinion to create the best forecasts and what are the best qualitative/quantitative methods to focus on?*

Gruetzemacher has recently proposed new methods aimed at forecasting transformative AI (Gruetzemacher 2019). Work toward this end - the development of novel methods for forecasting AI progress - is always welcome. Gruetzemacher proposes the notion of a holistic forecasting framework as well as an example of this for use in the context of AI. The only other example to meet the criteria of a holistic forecasting framework is that of Tetlock's (2017) full-inference-cycle tournaments which have received renewed attention for the purpose of AI forecasting (Gruetzemacher 2020). Both of these recent examples, and their suitability for the purpose of forecasting AI progress, suggest that there may be further value in pursuing the development of novel techniques. Separately, it is also likely useful to conduct studies to verify each of these proposed techniques as they have yet to be demonstrated in practice. Aggregation of indicators is also challenging, and an area of research that could be very useful, particularly if indicators could be aggregated into an a priori model of AI progress within a specified scope. More generally, work on a priori models is also something likely of value to the AI forecasting community.

---

[12] Interested readers can see https://aiimpacts.org/discontinuous-progress-in-history-an-update/.
[13] Significant work on this and similar topics is shared through only collaboration platforms, such as Google Docs.

### Long-term Forecasting

> *Q.8 How effective is long-term technological forecasting and how can we best validate near- and mid-term forecasts of AI progress?*

Little work exists regarding the effectiveness[14] of long-range forecasting[15]. Tetlock and Gardner (2016) suggest that a limit to geopolitical forecasts of roughly 5 years, yet Moore's law was a strong indicator of semiconductor progress for nearly 50 years and might be considered a successful long-term technology forecast. These examples illuminate something relevant to long-term forecasts but which has received little attention for near- and mid-term forecasts as well: the effectiveness of different methods for different types of forecasts (e.g. geopolitical and technological[16]). One such study only considered computers and not AI specifically (Mullins 2012)[17]. A valuable project would be to obtain the data from this study and evaluate it with a distinction between computing and AI. It would also be prudent to follow up with the sponsoring organization about any other work conducted since. Nagy et al. (2013) showed that extrapolation is effective for technology forecasting, however, there is no evidence that this applies to AI.

Long-term forecasts require substantial time to verify and thus it is difficult to determine the effectiveness of such forecasts. However, studies exploring the quality of forecasts for periods of five-to-ten years are recommended, particularly if they utilize individuals with a demonstrated aptitude for forecasting (e.g. superforecasters)[18]. Such studies may not yield quick returns, but could be very valuable for the community. Furthermore, another effort like that proposed by Mullins (2012), which was intended to obtain ~1,000 historical technological forecasts for comparison could be useful. An alternate approach would be the development of predictive models of different methods' or experts' forecasting accuracy (or a related metric) using the forecasting horizon (in years) as one of the inputs. While certainly challenging, due to the myriad of factors that can influence a forecast's accuracy, this approach does not necessarily require resolved long term forecasts to be useful, such that it could be applied to a wider variety of forecasting techniques.

The question of how to best validate forecasts - near- mid- or long-term - was found to be the most important and most feasible from the Delphi (Q8a; see Appendix C). For this reason we underscore the importance of this section, but also the importance of not just validating long-term forecasts, but near-to mid-term forecasts as well. Another open question is whether or not near-term progress can be used to validate the quality of mid-to long-term forecasts before these forecasts have resolved.

---

[14] Effectiveness was not defined in the Delphi process. We interpret effectiveness as being a combination of both accuracy and precision, but any study evaluating the effectiveness of forecasts should be careful to clearly define the term.

[15] This was not defined in the Delphi process, but for the purpose of this discussion we define: near-term forecasting as forecasts less than two years; mid-term forecasts as forecasts between two and five years; and long-term forecasts as forecasts beyond five years.

[16] It is possible that different types of questions are more forecastable for mid-to long-term forecasts, e.g. "will the United States be a nation in 10 years" may be more tractable to forecast than "will the United States President be Tom Cotton in 10 years." A study to evaluate the effectiveness of long-term forecasting relevant to these differing types of questions would be valuable for both AI forecasting efforts and broader forecasting efforts.

[17] Interested readers can also see Kott and Perconti (2018), and Muehlhauser (2017; 2019).

[18] It would be particularly useful to use a long-term forecasting study to calibrate a cohort of forecasters or superforecasters in order to use this cohort for future forecasts. It would be important to solicit participants with a high likelihood of continued participation after the full length of the study. It could also be useful to establish whether or not calibration on short-to mid-term forecasts was correlated with calibration on long-term forecasts after controlling for exogenous factors.

### 5.1.3 Dissemination and Miscellaneous Topics

#### Dissemination

> *Q.9    How do we utilize forecasts to inform decision makers and develop appropriate and measured initiatives/interventions?*

The dissemination of forecasts is a tricky but crucial issue; one common technique is scenario planning (Roper et al. 2011, Gregory and Duran 2001). Intuitive logics scenarios have been suggested as appropriate for this, but in the context of AI forecasts more complex scenario planning techniques may be required, such as scenario mapping techniques (Gruetzemacher 2019). It would be valuable to study the effects of disseminating technological forecasts using different techniques, such as these different forms of scenario planning techniques. Such a study would be valuable for forecasting AI progress, but also for the broader technological forecasting community. Further research could also focus on the role factors like prior exposure to the topic, the perceived intentions of the scenario presenter and the plausibility of presentation play in effectively conveying/disseminating AI forecasts. Moreover, as many forecasts are probabilistic, as this is a desirable quality, there is likely substantial value in reviewing known failure modes in communicating probabilistic information (Fischhoff 1994, Gigerenzer and Edwards 2003) as well as adapting communication strategies from fields like climate and natural disaster forecasting (Stephens et al 2012, Doyle et al 2014) to the context of AI.

#### Miscellaneous

The questions that fall into this category were not originally determined to fit neatly into one of the previous nine question clusters for which research suggestions have been included here. Because of this there is significant variance between the importance of the different questions in this category. Two questions have been identified from these six for discussion. First, the remaining four questions are first discussed, then the subsection ends with discussion of these two salient miscellaneous questions.

Most of the remaining questions could have arguably been included in one of the other clusters. One question which likely does not fit into existing clusters concerns a "minimum viable timeline" for radically transformative AI due to the catastrophic and existential risks associated with such extreme AI. Quantifying risks associated with such powerful AI systems is a problem of deep uncertainty, and this question attempts to raise an issue of decision making under deep uncertainty. Another question, which focuses on operationalizing group forecasting techniques, is generally a valuable area of research for both AI forecasting as well as all other applications of group forecasting techniques. Yet another question, concerning learning from existing long-range forecasting techniques, is likely well-suited to exploration through literature review and application. Such research may be worthwhile if it effectively extends the large body of existing work on the topic.

> *Q.10    How do we forecast the automatability of different types of unique human tasks?*

Future of work research is an important topic which is receiving substantial attention already involving both data based methods (Das et al. 2020; Martinez-Plumed et al. 2020) and expert elicitation (Duckworth et al. 2019). The data based techniques explored here only scratch the surface; efforts to obtain more datasets and to combine them with disparate data sources, either public or private, are worthwhile.

Existing work on this topic has suggested that it is possible to automate close to 50% of human jobs in coming decades (Frey and Osborne 2017), however, little work has explored the potential for extreme labor displacement from AI (Gruetzemacher et al. forthcoming). Models that can account for discontinuous progress in narrow domains (or more broadly) could be very useful for helping policy makers and organizations prepare for unforeseen scenarios. Thus, research on this topic can be very useful.

> *Q.11* *How effective are existing methods at technological forecasting (e.g. prediction markets, the Delphi technique, forecasting tournaments)?*

Long-term technological forecasts or AI forecasts are not the only type of forecast that it would be useful to validate. The validation of near- to mid-term technological forecasts, and AI forecasts specifically, would be very useful for comparison to explore the utility of different methodologies and the success rates in different subdomains of AI. A large-scale study of this would be most useful, but smaller scale studies of limited scope could also be very valuable to the community.

## 5.2 Methods-related Research Directions

A large portion of methods-related research suggestions have already been discussed in the preceding section. However, some topics did not emerge in the discussion of the research suggestions for the salient research questions. These are included in this section.

### 5.2.1 Statistical Methods Topics

Extrapolation is the simplest forecasting technique yet it remains one of the most valuable, even for the purpose of forecasting AI progress. The challenge lies not in extrapolating a trend from data, but from identifying an indicator with sufficient data that is also a signal of something important to decision makers. Thus, thinking critically about what may be a good indicator of AI progress is always valuable. This doesn't necessarily require focus and dedicated time, but rather motivation and genuine interest in understanding AI progress. A valuable project would be to create a git repository where data for all proposed indicators can be aggregated. This is similar to the proposed AIcollaboratory (Martinez-Plumed et al. 2020), but it would also be useful to include social indicators and other indicators beyond benchmarks or measures such as computational resources (Amodei and Hernandez 2020) or algorithmic efficiency (Hernandez and Brown 2020).

Indicator selection is particularly challenging, and it is important for interested parties to be cognizant of lessons from existing work. For one, it is often easier to extrapolate from benchmarks that are far removed from a specific task or a proxy for human performance. For example, log-likelihood loss may scale continuously with computational resources yet we do not have a good understanding of what this indicator implies for performance on future tasks or its relation to human performance (Kaplan et al. 2020). However, extrapolation of indicators at the task-level is frequently not smooth in the manner that log-likelihood loss is (Brundage and Clark 2017). It is also important to note that the notion of "human-level" performance on a certain task can evolve over time because benchmarks and metrics are often poor proxies for the performance of a human on complex tasks such as visual recognition or natural language understanding.

## 5.1.2 Judgmental Methods Topics

Despite the extensive body of literature on technological forecasting, there is still little work comparing the performance of the best performing judgmental forecasting techniques specifically in the context of AI forecasting. While such comparisons have been conducted for judgmental techniques more generally (Green et al. 2015), it remains unclear whether these results are indicative of the performance of these techniques in technological forecasting applications. Here, we are interested in investigating the effectiveness of different techniques for not one but two different applications: 1) technological forecasting and 2) AI forecasting. One simple and straightforward project on this topic would be conducting a Delphi study involving PhD students studying AI using the same forecasting targets as those posted to ai.metaculus. Similarly, one could conduct a survey and structured interviews with PhD students studying AI using the same forecasting targets as those posted to ai.metaculus (on or near the closing date for the forecasts). Analysis of results from experiments like those proposed could be of immediate value to researchers who are actively using expert elicitation with experts. We do not recommend those inexperienced with expert elicitation attempt working with experts due to the risk of fatigue and future nonparticipation. Because the body of experts is so small, and their opinion may play a crucial role in future forecasts, we perceive this to be a serious risk[19].

      Similarly, work could be conducted to evaluate and improve scenario analysis techniques applied to forecasting AI progress in a variety of applications, e.g. forecasting target generation. For example, with PhD students studying AI, we suggest using an established method to conduct scenario analysis on near-term plausible forecasting targets such as facial recognition technology, autonomous vehicles and lethal autonomous weapons. This process could be performed for multiple groups of students, focusing on a one or two year time horizon and meticulously documenting the facilitation process. The results could be evaluated to identify how the technique can be improved.

      One particularly interesting methods-related topic in need of further exploration is the use of tools like Foretold[20], Elicit[21], or Metaculus's probability interface for eliciting probability distributions instead of point estimates or probability quantiles. Simple experiments could be devised to evaluate the impact of using this technique for eliciting distributional forecasts. Results from such studies would likely be valuable beyond the AI forecasting community and would be of interest to the broader forecasting community. Moreover, it is likely that novel techniques would be necessary for aggregation and analysis of results from such studies, which could lead to further novel work relevant to forecasting beyond just forecasting AI progress and technological forecasting.

## 5.2.3 Hybrid Methods and Miscellaneous Topics

Little work exists on hybrid methods, but Gruetzemacher (2019a) is a good starting point for interested researchers. As noted earlier, work on novel methods is welcome, and this is particularly true for the development of hybrid techniques. This includes techniques which would meet the criteria of a holistic forecasting framework, such as that of Tetlock (2017), as well as hybrid techniques which aren't cyclic in nature.

---

[19] Anyone planning expert elicitation with AI experts should seek guidance of those with expert elicitation experience in the context of forecasting AI progress.
[20] www.foretold.io.
[21] https://elicit.ought.org/.

In-depth analysis of specific questions casts a very broad net for possible research topics, and we hope that many readers are able to do better than us. However, examples include: 1) What indicators or milestones could be expected to precede discontinuous progress toward radically transformative AI? 2) Would a complete solution to the problem of meta-learning, in combination with a suite of powerful, specialized deep learning subsystems, be enough to enable some form of radically transformative AI?

Perhaps most significant of the miscellaneous methodological topics is the importance of literature reviews. *For a large portion of the research suggestions throughout this agenda, the foremost priority should be a survey of the existing literature.* Because literature reviews are low hanging fruit that can be accomplished with minimal resources and prior knowledge, we suggest that readers new to this topic and interested in contributing first attempt a literature review[22].

Participants in the Delphi process may have overlooked some miscellaneous topics. We feel that the definition of terms used in AI forecasting efforts is crucial, and that ontologies could be useful in this context. Work from Lagerros and Goldhaber (2019a) could be a good starting point for extending further work. It is also useful to explore further efforts described by Lagerros and Goldhaber (2019b) regarding resolution councils for difficult to resolve forecasting targets (common in AI forecasting due to the complex nature of targets which yield strong signal).

# 6. Discussion & Limitations

## 6.1 Discussion

The research suggestions compiled in the previous sections are not meant to be comprehensive or to suggest that any of the questions or methods discussed be ignored or deprioritized. Rather, they are intended to provide examples for how to translate the results of the Delphi process into concrete research proposals as well as to provide a starting point for researchers from adjacent fields and junior researchers. As researchers working in AI forecasting, we are all similarly familiar with the body of existing work as the experts who participated in the Delphi process, which is why numerous concrete suggestions for the questions discussed in Section 5 involved simply beginning with a literature review. We were pleased that the experts also scored this to be of above average importance, and more significantly, to be the most feasible technique. Consequently, *we believe that literature reviews should receive the highest priority by motivated researchers*[23].

This was an ambitious project, not because we sought to use the Delphi to identify salient research topics for a certain research domain, but because we chose to do so in a manner that generated a publishable research agenda which addressed a significant gap in the existing body of literature rather than just a paper documenting the process (Kellum et al. 2008; Dahmen et al. 2013; Burt et al. 2009; Kaiser and Deaver 2013; Dimmit et al. 2005; Gordon & Barry 2006; Beddoe et al. 2016). While we learned a lot about how the process can be improved for future elicitations similar in nature, we are also pleased with the results and hope that this research agenda is able to make a positive impact on the future of research concerning AI forecasting.

---

[22] The authors also ask anyone who completes a related literature review to please contact us so that we can compile a list of relevant literature reviews.

[23] Literature reviews could require less prior context than other projects to be successful (as context is acquired in the process), so they may be well-suited for non-experts.

## 6.2 Limitations

There are a number of limitations of this study, several of which we briefly describe here. The first major challenge involved the lack of scores from some participants for many of the questions and methods during the second round of the Delphi. This increased uncertainty because it is impossible to know the intentions of the experts regarding their failure to answer certain questions. Because a large number of participants all exhibited similar behavior, the problem was more pronounced than anticipated. However, it is impossible to know the counterfactual, and we feel that the 80% 2nd round response rate from 1st round participants was worth the challenges that these missing values posed. Moreover, efforts were made to address issues caused by the missing values, but we ultimately found these efforts to be insufficient.

Another significant challenge was present in the numerous difficult decisions necessary to best present the data and results of this study. The study itself finds the dissemination of forecasting results to be a topic of high importance and above average feasibility. We also believe dissemination to be important for this study, but we found it to be more challenging than the above average feasibility scores from the Delphi would suggest. Consequently, we have done our best to present the material in a straightforward and objective manner that can also be easily digested and referenced for promoting future research. To these ends we have included four appendices, one of which contains further details about the Delphi process and another three which contain more detailed discussion of the Delphi results.

Another limitation was due to the limited amount of researchers currently working on forecasting AI and our limited rate of participation; the results of the Delphi were likely affected by self-sampling bias. We found some evidence for this hypothesis, as practitioners seem to be overrepresented in the respondents[24], compared to researchers. While these issues limit the conclusions that can be rigorously drawn from our data, this is not too much of a problem given the exploratory scope of this document. Even if there was strong self-sampling bias, the document still represents the opinions of a significant fraction of the relevant research community and thus provides a strong starting point for researchers interested in engaging with the field.

# 7. Conclusion

A Delphi study was conducted involving experts with experience related to forecasting AI progress in order to produce a research agenda on this topic. AI is a general purpose technology poised to transform business and society over the coming decades, and forecasting progress in this field is critical for informing policy makers and decision makers so that the impacts are managed effectively and in a manner that is beneficial to mankind. Despite being such an important endeavor, there is no document in the existing literature framing this problem and motivating rigorous academic work on the subject. Our study addressed this gap in the literature and went further to elicit experts regarding the paths to prioritize for researchers interested in working on forecasting AI progress.

The results represented a wide range of important questions and methods for those interested in the topic to focus in future work. The results were complex to present, and there are many issues that did not receive due attention in this summary. All of the questions and methods described in here are worth pursuing because forecasting AI progress poses challenges more daunting than those posed by other technologies.

---

[24] See Appendix A, Table AI.

AI is a very powerful technology, and perhaps more dangerous than any technology that has come before it. It is of the utmost importance to ensure that all efforts to mitigate risks posed by AI are taken, and in order to do this it is necessary to correctly anticipate the technologies and the timelines for their arrival. We hope that this effort to objectively determine the most important issues for this crucial problem can be useful to both researchers and practitioners, so that it may truly have a positive impact on the future decisions about AI that will undoubtedly carry with them consequences that could last for many generations.

# Acknowledgements

We would like to thank the anonymous Delphi participants and to acknowledge the AI Safety Research Program. We also thank Tamay Besiroglu, Jose Hernandez-Orallo, Miles Brundage and Ozzie Gooen for their comments on drafts of this manuscript.

# Appendix A: The Delphi Process and 2nd Round Results

The experts we invited for the Delphi study had previous experience in either economics, technological forecasting or AI forecasting or the use of foresight techniques for AI. These experts were representative of the academic, government, industry and nonprofit sectors. Of the 15 of 32 experts who responded to the survey, four were from industry, two were from government, four were from nonprofits and five were from academia. Of the ten who had published on the topic, there was a mean and median of 5,977.6 and 1,195 citations, respectively[25], with a mean and median h-index of 22.3 and 11, respectively. Of the experts who did not respond seven were from academia, none were from government, four were from industry and six were from nonprofits. A breakdown of respondents and nonrespondents can be seen in Table I. All nonrespondents had published on the topic and had a mean and median of 7,455.3 and 2088 citations, respectively, with a mean and median h-index of 21.1 and 12 respectively.

**Table A1: Responses and Non Responses by Employer**

|  | Response | No Response |
| --- | --- | --- |
| Academics | 5 | 7 |
| Government | 2 | 0 |
| Industry | 4 | 4 |
| Nonprofit | 4 | 6 |
| Total | 15 | 17 |

The primary difference between respondents and nonrespondents is that five of the respondents have not published on the topic while all nonrespondents have published on the topic. All of these respondents were involved directly in the development of forecasting platforms (e.g. prediction markets) and were actively involved in efforts to forecast AI progress. It is also of significance that the nonrespondents were more academically accomplished in terms of citations, but in terms of h-index the respondents and nonrespondents were more comparable. Of nonrespondents, it is significant that six were not known to be actively working on projects related to forecasting AI progress. Of the AI foresight experts, one responded while the other did not. We also find it important that both of the invited participants from governments (i.e. the US and the EU) chose to participate.

Due to the scarcity of experts in this domain who could contribute to this study, only 32 experts were identified for soliciting. Consequently, it can be expected to have a marginally smaller participation rate from the more successful researchers, which would help to explain the discrepancy in number of citations. However, despite the differences between the respondents and the invited experts, we believe the sample to be generally representative of the broader group of invitees. An analysis of the experts who did not choose to participate, given their bodies of existing work, does not appear to suggest that the broader trends drawn from the results presented in this section would be different given other random samples.

---
[25] Based on Google Scholar, accessed February 15th, 2020.

Invitations for the questionnaire were sent to invitees via email. If participants did not respond in the first five days they were sent a reminder email. Six participants responded initially, and nine more responded following the reminder email.

Delphi Questionnaire
1. Do you feel that forecasting AI progress is, or could be, a well defined research topic? Why?
2. What questions should researchers who work on forecasting AI progress prioritize?
3. What methods or techniques should researchers use/prioritize to answer these questions?
4. Are there any topics relevant to forecasting AI progress that you feel are important but neglected? Why?

Following the Delphi questionnaire, a novel step was introduced to enable discussion among the participants. This step first required the aggregation of all of the participants' answers to the four questions of the questionnaire. Following aggregation of the answers, the results were placed in a Google Doc. This included an overall summary as well as hierarchical lists of the most salient questions and methods. The questions and methods had been clustered in order to reduce redundancy. In this summary document, participants were only given the ability to comment and suggest. The participants were then emailed and instructed to make any suggestions for changes anonymously. They were also instructed to have anonymous discussions in the comments of the document if there was any disagreement. This was described to participants as an optional portion of the first round of the Delphi in order to reduce attrition. Only two minor comments were registered during this period[26].

After the discussion period, a Google Sheet that was created to capture the hierarchical layout of the list of questions and methods from the previous rounds. In order to increase participation, respondents were not required to give ratings on all items and partial answers were accepted. Participants were then instructed to rate the questions and methods on a scale from 1 to 5 along the axes of importance and feasibility. 5/15 participants completed the rating within the first five days and seven additional responses were obtained after sending a reminder email.

The full results of the Delphi are reported in Tables II and IV. The hierarchical form of the questions reported to participants after the first round of the Delphi is presented in Appendix B. This hierarchical form was derived from the facilitators' assessment of the question clusters' relative importance following the first round of the process. The remainder of this section follows the structure of the questions depicted in Appendix B, focusing most on the questions that received strong interest in both the 1st and 2nd rounds of the Delphi process.

---

[26] The summary and lists, as well as the recorded comments can be found at this following link: https://tinyurl.com/AI-Forecasting-Delphi-Round-1. The content is also included in Appendix B.

# Appendix B: 1st Round Delphi Results (Questions and Methods)

The content here is the results of the 1st round of the Delphi that was delivered to the participants during the discussion phase of the Delphi. First, the summary of the results is depicted. The questions and methods are then shown, hierarchically, to reflect the structure and importance that the facilitators agreed upon.

## B.1   1st Round Results (Summary)

There was general consensus that forecasting AI progress is or, with appropriate effort, could be a well-defined research field. Regarding questions to prioritize, the most common responses were broadly about the **AI production function:** taking a detailed quantitative look at **inputs** and a hard, critical look at **outputs/measures of progress.**

Three primary perspectives emerge on the methods that can be used to forecast AI progress:
- Statistical modeling using indicators or metrics for measuring AI progress (~60%)
- Judgmental forecasting techniques for exploring plausible paths forward and for eliciting probabilistic forecasts (~25%)
- Hybrid methods, which use elements of both statistical and judgmental forecasting techniques (~15%)

Regarding neglected topics, a few significant additions were made to the questions list, and others reinforced some of the salient questions/topics identified in the second prompt.
    We have attempted to organize the responses about questions of interest and most valuable methods below. We have taken the unique neglected topics, formulated them as questions and added them to the section below:

## B.2   1st Round Results (Questions)

The results shown here are structured based on the results from the 1st round of the Delphi process. Repeated questions were aggregated and similar questions were clustered with consensus from the four facilitators. The clusters (represented by top-level bullet points in the following list of results) were then first ordered by the amount of answers that related to them and then slightly rearranged by the facilitators to better reflect relationships between the clusters. Eight questions could not be clustered, all but one of which are included at the end of the list. No questions were excluded that were included from the respondents, but the majority of questions were paraphrased (through clustering and summarization) to represent them in as few words as possible. Thus, nuances were not included in the reported results.

1. What are the most important forecasting targets?
    - How do we define qualitative and quantitative measures of progress toward forecasting targets?

- How can we decompose abstract AI technologies into more easily forecastable targets?
- What questions/targets matter for practical, near-term decision making?
2. What are the implications of timelines?
    - Should we focus on capabilities or the impact of AI systems?
    - How can forecasts be applied to identifying and mitigating risks?
3. How do we best evaluate overall AI progress?

4. What are the most useful indicators (e.g. compute, talent, investment/resources, economic impact, benchmark performance)?
    - What performance metrics are relevant and most effective?
        - How do we assess the quality of a metric/benchmark's signal?
        - Are existing (SOTA) benchmarks relevant or useful (i.e. strong signal)?
        - Should we focus on tasks or abilities for measuring and forecasting AI progress?
        - How would we develop a broader discipline for measuring and assessing progress in AI (like psychometrics)?
        - How do we best analyze/measure AI systems' abilities to generalize, understand language and perform common sense reasoning?

5. How can we model AI progress?
    - What are the best methods for modeling given the correct variables?
    - Why is progress faster in some metrics than others?
    - Can independent variables be used to model AI progress effectively model progress in other fields/research domains?

6. What are the most probable AI development scenarios?
    - How do we identify the most plausible paths for a variety of transformative AI technologies/systems?
    - What will be the new applications/services made possible by new AI technologies?
    - What impact does NLP have on AI capabilities?

7. How do we produce the best forecasts?
    - How do we aggregate and report metrics?
    - What are/how do we develop the best qualitative/quantitative a priori models?

8. How effective can long term forecasting of AI progress be?
    - How do we best validate forecasts of AI progress: historical data/near-term progress?

9. How do we utilize forecasts to inform decision makers and to develop appropriate and measured initiatives/interventions?
    - Who are the relevant stakeholders/audiences for forecasts and how do we best report forecasts to each?
    - What are information hazards related to AI forecasts and how do we best make decisions about how to guard and disseminate forecasting data?
    - What can we learn from historical examples of policy making?

10. How can we improve/make more useful conventions regarding forecasting questions and answers?
11. How do we forecast the automatability of different types of unique human tasks?
12. How can we collect data measuring human performance that can easily be compared to machine performance (e.g. next word prediction log loss)?
13. Can we identify a minimum viable timeline (e.g. 10% of strong AI) for use by stakeholders and decision makers?
14. What can we learn from existing long-range forecasting techniques (e.g. clionomics, K-wave theory, S-curves)?
15. How do we best operationalise group forecasting efforts?
16. How effective are existing methods at forecasting technology (e.g. prediction markets, the Delphi)?

## B.3   1st Round Results (Methods)

The 1st round results were collected into three primary classes. These classes were included because they are reasonable for differentiating between the different forecasting techniques.

A. Statistical forecasting techniques:
   a. Statistical modeling
      i. Extrapolation
      ii. Bayesian methods
   b. Benchmarks & metrics
      i. Aggregating into metrics for human comparison
      ii. Item response theory
   c. Data science (e.g. tech mining, bibliometrics, scientometrics)
      i. Theoretical models
   d. Machine learning modeling
   e. Simulation
B. Judgmental forecasting techniques:
   a. Simulation & role-play games
   b. Scenario analysis
   c. Blue-team/red-team
   d. Expert elicitation:
      i. Delphi
      ii. Expert adjustment
   e. Prediction markets
   f. Forecasting tournaments
   g. Calibration training
   h. Aggregation of expert opinion
   i. Immersive observation of AI labs
   j. Identifying clear and effective forecasting targets
   k. Conceptual progress acceleration survey (using pairwise comparisons)
C. Hybrid methods (i.e. statistical and judgmental)
D. Other:
   a. Probabilistic reasoning (e.g. the Doomsday argument)
   b. In-depth analysis of specific questions
   c. Literature review

# Appendix C: Top Questions Results

Appendix B and Appendix C have reported the raw results from the first and second rounds of the Delphi study, respectively. This section discusses these results, exploring them as the groups of questions clusters that they were organized in when aggregated following the first round of the Delphi. The mean importance and mean feasibility score for each specific question are included in parentheses following each question, respectively. The questions with the ten[27] highest importance scores are depicted in **bold** and the top three highest feasibility scores are *italicized*[28]. For topics we believe to be important foci for future work we use the words priority or prioritize and we underline these words to signal to the reader these recommendations. However, we do not make many recommendations because of the uncertainty due to selective scoring, and we only make recommendations that we are confident in given the data.

## C1 Meta-level Topics

The questions in this section all involve meta-level topics about AI forecasting, specifically, what forecasting targets are most relevant for decision makers, the implications of forecasts, and how to evaluate overall progress in AI. The first two topics appear as clusters of related questions, while the final question relates to "overall" progress.

### Forecasting Targets

1. What are the most important forecasting targets?[29]
    a. How do we define qualitative and quantitative measures of progress toward forecasting targets? (4.00, 3.65)
    b. **How can we decompose abstract AI technologies into more easily forecastable targets? (4.11, 3.64)**
    c. **What questions/targets matter for practical, near-term decision making? (4.48, 3.83)**

Forecasting targets are critical for judgmental forecasting techniques and their significance is reflected in the experts' scoring of this cluster. All of these questions scored in the top 50% by importance, with Q1b scoring in the top quartile and Q1c scoring in the top 10%; all of the feasibility scores fell within top 35% highest scored questions. However, the significance of these questions cannot all be attributed to the importance of correctly specifying what is to be forecast when using judgmental techniques because Q1c includes questions as well as targets, and this is the 3rd highest scoring question of all. It is also interesting that the two top 10 questions, Q1b and Q1c, are scored well above average on feasibility, and could be interpreted to suggest that these topics are some of the most promising directions for future work.

---

[27] This is actually 11 questions because there is a tie for 10th.
[28] This is actually 4 questions because there is a tie for 3rd.
[29] Every question in the outline depicted in Appendix B was included in the scoring during the 2nd round with the exception of the first question, i.e. the first question in the first cluster here. An error was made when creating the elicitation for the scores which resulted in this question being left out. It is unfortunate this transcription error led to the parent question for this group being excluded from the 2nd round scoring elicitation because it is unclear how it would have been scored.

### AI Timelines

2. What are the implications of timelines? (3.68, 3.14)
    a. Should we focus on capabilities or the impact of AI systems? (3.42, 2.69)
    b. How can forecasts be applied to identifying and mitigating risks? (4.04, 2.88)

### Evaluating Progress

3. How do we best evaluate overall AI progress? (3.93, 3.0)

While the implications of timelines are an important topic, they are not a topic that was prioritized by the experts. The importance scores range from a full standard deviation below the mean to the 3rd quartile, and the feasibility scores are all below the mean, with 2a falling over a full standard deviation below the mean for feasibility as well as importance. The importance of questions in this cluster could be interpreted to suggest that the most significant implications of timelines - in the context of AI forecasting - are conclusions concerning risk mitigation (Q2b), but the below average feasibility scores suggest that practically using forecasts to mitigate risks can be challenging.

How to best evaluate overall AI progress is also a topic of significant importance, but one which is challenging to address as reflected by the low feasibility score. This single question is not truly a cluster, but it did fit well with the rest of the group because it is a very general, high-level question. To answer this question presents a broad and daunting challenge, but one that is worth pursuing because it may yield great benefits. It is interesting that this group of topics (i.e. Q1-Q3) receives as much attention as it does because other technologies that commonly receive attention in the forecasting literature do not seem to face challenges this general in nature. This could have something to do with the fact that AI is considered a general purpose technology (Brynjolfsson et al. 2017), and little work has been done on forecasting progress toward other general purpose technologies in a broad sense (Gruetzemacher et al. forthcoming).

## C2 Forecasting Methods-related Topics

This section covers five question clusters all relating to forecasting methods, covering specific topics like: 4)[30] identifying useful indicators, 5) methods for best modeling progress, 6) identifying the most probable development scenarios, 7) supplemental methods to improve forecasts and 8) the value of long-term forecasts.

### Indicators and Metrics

4. What are the most useful indicators (e.g. computation, talent, investment/resources, economic impact, benchmark performance)? (4.10, 3.89)
    a. **What performance metrics are relevant and most effective? (4.26, 3.54)**
        i. **How do we assess the quality of a metric/benchmark's signal? (4.2, 3.44)**
        ii. Are existing (state-of-the-art, SOTA) benchmarks relevant or useful (i.e. strong signal)? (3.87, 3.78)

---

[30] The numbering followed for the clusters is consistent with the numbering of the entire results included in Appendix B.

      iii. Should we focus on tasks or abilities for measuring and forecasting AI progress? (3.22, 2.38)
      iv. How would we develop a broader discipline for measuring and assessing progress in AI (like psychometrics)? (3.00, 3.09)
      v. How do we best analyze/measure AI systems' abilities to generalize, understand language and perform common sense reasoning? (3.77, 2.77)

Metrics and various measures of performance or progress are very important to experts based on the importance scores, with over half of the questions scoring above average. Q4, Q4a and Q4a1 all score strongly on importance as well as above average on feasibility, with the latter two in the top ten. Q4a2 also scores above average on both importance and feasibility. The remaining three questions each score below average for importance and well below average for feasibility, however we do not necessarily interpret this all to diminish the significance implicated from the previous questions. For example, Q6a.iv is similar to the notion of "aggregating into metrics for human comparison" (M-Ab.ii), which was found to be the 3rd most important method. Consequently, this question may have received diminished importance because it was imprecisely stated.

## Modeling AI Progress

5. How can we model AI progress? (3.76, 3.21)
   a. What are the best methods for modeling given the correct variables? (3.29, 3.33)
   b. Why is progress faster in some metrics than others? (3.40, 3.56)
   c. Can independent variables be used to model AI progress effectively model progress in other fields/research domains? (3.31, 3.12)

This cluster can be summarized by its focus on modelling progress in AI. Notably, all questions scored below average for importance and the feasibility scores all fell within the interquartile range. Overall, this cluster could be interpreted to suggest a lack of consensus among experts regarding the use of models for AI forecasting. Modelling is also a broad term, and it is noticeable that the more specific questions score significantly lower than the sole general question (Q5).

## Concrete Scenarios

6. **What are the most probable AI development scenarios? (4.11, 2.86)**
   a. **How do we identify the most plausible paths for a variety of transformative AI technologies/systems? (4.20, 2.83)**
   b. What will be the new applications/services made possible by new AI technologies? (3.41, 2.49)
   c. What impact does NLP have on AI capabilities? (3.34, 3.21)

This sixth cluster focuses on scenarios for AI development. Some questions in this cluster, such as Q6 and Q6a, are considered to be very important as they score in the top 10 most important questions. However, other questions, such as Q6b and Q6c, are much less significant scoring more than a standard deviation below the mean importance score. Notably, all of the questions score below average on feasibility. It is interesting that despite tremendous recent progress in natural language processing (NLP; Raffel et al. 2019;

Radford et al. 2019), the implications of this progress are not scored highly among forecasters, but this is not necessarily surprising considering the emerging trend for more general questions to be scored higher. This dichotomy is illustrated in experts' ratings: while the more general questions' importance were the 5th and 10th most important questions, the more specific questions both fell in the bottom quartile. It is interesting to see two questions scoring very highly on importance but also less feasible. This might suggest a level of consensus among experts that scenario analysis is important yet challenging.

## Improving Forecasting Efforts

7. How do we produce the best forecasts? (4.03, 3.06)
    a. How do we aggregate and report metrics? (4.09, *4.31*)[31]
    b. What are/how do we develop the best qualitative/quantitative a priori models? (3.93, 2.78)

The next cluster contains only three questions, loosely related to generally improving forecasting efforts around AI. These questions all score highly for importance, falling in the third quartile, and are mixed regarding feasibility. Q7 and Q7b both score below average for feasibility, but Q7a scores the highest of all questions with respect to feasibility. These results are perhaps unsurprising given the empirical success of aggregation methods in forecasting (Ungar et al. 2012) and the seeming amenability of these methods to theoretical analysis (Satopää et al. 2016). Because this is also scored as important, we interpret it to suggest that work which draws from this body of existing work to develop AI-specific techniques for the aggregation and reporting of metrics should be prioritized.

## Long Term Forecasting

8. How effective can long term forecasting of AI progress be? (4.10, 2.13)
    a. **How do we best validate forecasts of AI progress: historical data/near-term progress? (4.62, 3.84)**

This final cluster in the group is concerned with the general effectiveness of long term forecasting, as well as the validation of forecasts about AI progress. Both of these questions are central to forecasting AI progress: as many of the transformative effects of AI might still be relatively far in the future, it is vital to know how valid AI-specific forecasts are over near-, mid- and long-term time horizons. Furthermore, feedback about previous forecasts' outcomes plays an important role for individual forecasters (Harvey 2001, Fischhoff 2001), as well as the selection of forecasting methods (Armstrong 2001). However, long-term forecasts obviously require a long time to evaluate and most existing long-term forecasts were not stated precisely (Mullins, 2018), thus, it has been difficult to evaluate them rigorously enough to draw meaningful conclusions (Muehlhauser, 2019).

It is unsurprising that each of these questions score high on importance. Regarding long-term forecasting, the median importance for this question was 4, with 7 of 9 forecasters scoring it at 4 or higher. This suggests a strong consensus about its importance. However, while desirable, it was scored to be the least feasible of all questions. Perhaps more significantly is Q8a which is the highest scored question of the study. Moreover, this question scored a standard deviation above the mean with respect to feasibility. *Q8a is obviously an important question, and the strength of its scores for both importance and*

---

[31] This is underlined because it was scored the most feasible question by a relatively large margin.

*feasibility clearly indicate that this should be considered the highest <u>priority</u> research question*[32].

## C3 Dissemination and Miscellaneous Topics

This section includes all of the remaining questions. These only form one significant cluster which did not fit well with either of the other two groups. This cluster is presented first below. The remaining questions did not fit well into any of the previous clusters or groups, so they are presented here independently.

9. **How do we utilize forecasts to inform decision makers and to develop appropriate and measured initiatives/interventions? (4.54, 3.34)**
   a. Who are the relevant stakeholders/audiences for forecasts and how do we best report forecasts to each? (3.78, *4.24*)
   b. **What are information hazards[33] related to AI forecasts and how do we best make decisions about how to guard and disseminate forecasting data? (4.17, 3.99)**
   c. What can we learn from historical examples of policy making? (3.51, 3.68)

This last question cluster did not fit in either of the previous two groups because it concerns topics related to ensuring the safe and effective dissemination of forecasts to decision makers. This is a crucial cluster because improved decision making is the ultimate goal of forecasting (Armstrong 2001), but accurate and precise forecasts about complex technological issues are not useful if they are misunderstood or misinterpreted by decision makers. Two of the questions (Q9 and Q9b) in this cluster scored among the top 10 on importance, and, notably each of these also scored above average for feasibility. The other two questions (Q9a and Q9c) scored below average on importance but above average on feasibility. The two top scoring questions make this one of the more significant question clusters and can be interpreted to suggest that the participants place a very high value on the need to carefully report the results of forecasts pertaining to AI.

10. How can we improve/make more useful conventions regarding forecasting questions and answers? (3.3, 3.51)
11. **How do we forecast the automatability of different types of unique human tasks? (4.12, 3.36)**
12. How can we collect data measuring human performance that can easily be compared to machine performance (e.g. next word prediction log loss)? (3.52, 4.15)
13. Can we identify a minimum viable timeline (e.g. 10% of strong AI) for use by stakeholders and decision makers? (3.0, 2.5)

---

[32] As a way to choose the best methods and forecasters, we try to validate the predictions they make. However, there are not many data points to evaluate methods within AI since not that many proper forecasts have been carried out. To try to solve this, we rely on how methods have performed in technological forecasting and long-term forecasting which are the two areas with the most in common with AI forecasting. Increasing the data available for validation is a common concern in the field (Grace et al., 2018). There have also been attempts at covering the literature to find as many previous forecasts as possible and try to score them to get more information on what works well (Muehlhauser, 2016). However, there's still room for a more detailed review and we need more rounds of surveys to be able to check how forecasts work for longer horizons.

[33] See Bostrom 2011.

14. What can we learn from existing long-range forecasting techniques (e.g. clionomics, K-wave theory, S-curves)? (3.71, *4.29*)
15. How do we best operationalise group forecasting efforts? (3.69, 3.62)
16. **How effective are existing methods at forecasting technology (e.g. prediction markets, the Delphi)? (4.14, 4.02)**

These final 7 questions did not fit neatly into any of the previous clusters. However, because no questions were removed, they are reported included. It is interesting that two of the top 10 questions (Q11 and Q16) did not fall into any previous group or cluster. It is also interesting that the remaining questions (Q10, Q12, Q13, Q14 and Q15) were all scored to be of less than average importance. Q11 is a significant topic that has likely received more attention than any others included here because of its economic implications; research on this topic is widely referred to as the study of *future of work*. Q16 is also an important topic that is has oddly been understudied in the past; one possible reason could be the difficulty of finding true domain experts to participate in forecasts related to the expertise (Rowe and Wright 2001). However, forecasters participating in this elicitation scored it to be in the top five most feasible questions, leading us to conclude that this is a leading topic to prioritize. It is also worth noting that, while scoring below average on importance, two more of the top five most feasible questions (Q12 and Q14) are contained in this group. These questions could be worthwhile to pursue given the consensus around their feasibility among experts.

# Appendix D: Methods for AI Forecasting Results

Appendix B and Appendix C have reported the raw results from the first and second rounds of the Delphi study, respectively. This section discusses these results, exploring them as the different classes of methods in the order that they were organized when aggregated following the first round of the Delphi. Three methods only received 2 responses. These methods are identified where the scores are reported. All other methods received 5 or more responses. The top third of methods by importance (9 of 29) are marked in bold, while the top 3 feasibility scores are italicized[34][35].

## D1 Statistical Methods

Statistical methods refer to forecasting techniques that approach forecasts in a systematic way by taking either empirical or elicited data and using it as the input for a statistical model. Little published work exists that has attempted to develop rigorous statistical models for modeling AI progress, although such efforts have been outlined (Brundage 2016) and variables for such models have been proposed (Martinez-Plumed et al. 2018). Consequently, it is encouraging to see the experts' substantial interest in and perceived importance of these methods.

- A. Statistical forecasting techniques: (3.50, 3.0; <u>only 2 responses</u>)
    - a. **Statistical modeling (3.90, 3.9)**
        - i. **Extrapolation (4.07, *4.4*)**
        - ii. **Bayesian methods (3.88, 4.05)**
    - b. Benchmarks & metrics (3.82, 4.08)
        - i. **Aggregating into metrics for human comparison (4.20, 3.60)**
        - ii. Item response theory[36] (3.00, 4.5; <u>only 2 responses</u>)
    - c. Data science (e.g. tech mining, bibliometrics, scientometrics) (3.67, 4.0)
        - i. **Theoretical models (3.98, 3.3)**
    - d. Machine learning modeling (3.83, 3.67)
    - e. Simulation (3.59, 3.7)

The appearance of 5 of the top 9 highest scoring methods underscores the importance of this class of techniques; we assume the poor response rate for the class led to its below average importance score. Statistical modeling and each of the methods associated with it all scored among the highest third of methods based on importance. Moreover, these methods were all above average with respect to feasibility, with extrapolation being in the top 3 most feasible methods. Extrapolation is widely used and very successful in many applications, including for Russell and Norvig's (1995) correct prediction of DeepBlue's major milestone in 1997. Consequently, it is no surprise that experts scored it highly, and this leads us to conclude that it should be considered one of the most powerful techniques for forecasting AI progress. However, we note that, while very feasible, the challenge for extrapolation typically lies in identifying the appropriate indicators with strong signal of true progress in a subdomain or toward a specific objective.

---

[34] This is actually 4 methods because there is a tie for 3rd.
[35] Methods that only received two answers are excluded.
[36] A mathematical family of models that can be used to describe the nature of AI systems' abilities.

Benchmarks and metrics were overall identified to be of above average importance and feasibility. Item response theory only received two responses, so it was likely not well understood by many respondents[37]. However, aggregating into metrics for human comparison (M-Ab.ii) was found to be the 3rd most important method[38] but also scored slightly below average on feasibility. It is interesting that this scored so high on importance because the question which was scored as the least important (Q4a.iv) seems to simply suggest that a separate field of study is necessary to address this same issue. Consequently, it seems that semantics could have led to some very important topics not being recognized as such. Thus, it is important for those truly interested in contributions to this emerging research area to consider all of the questions and methods described herein.

Three of the four remaining methods in this class - those involving data science, theoretical models and machine learning - each scored at or above average for importance, with theoretical models scoring among the top third. While data science scored above average for feasibility, the other two were in the first and second quartile. Given the widespread use of data science and tech mining for technological forecasting applications, it is unexpected that this was not scored to be more important by the experts. The use of machine learning models for forecasting is relatively new, and consequently it is likely a useful topic to explore further[39]. Despite being among the best methods for forecasting catastrophic risks (Beard et al. 2020), simulation alone scored slightly below average for both importance and feasibility. This is likely due to the fact that data does not exist to create world models for forecasting AI progress like it does for climate change.

## D2 Judgmental

Judgmental forecasting techniques are not as widely used as statistical forecasting techniques in practice, however, for the purpose of technology forecasting, and AI forecasting in particular, they offer some unique advantages. The list of methods that were mentioned by experts are organized in the outline below. The remainder of this section analyzes these techniques more closely, considering the importance and feasibility scores and exploring concrete research suggestions for some.

   B. Judgmental forecasting techniques: (3.00, 3.50; <u>only 2 responses</u>)
      a. Simulation & role-play games (2.91, 4.38)
      b. Scenario analysis (3.59, 4.08)
      c. <u>Blue-team/red-team (3.80, *4.4*)</u>
      d. Expert elicitation: (3.38, 3.75)
         i. Delphi (3.69, 4.29)
         ii. Expert adjustment (3.80, *4.6*)
      e. Prediction markets (3.18, 3.71)
      f. Forecasting tournaments (2.99, 3.93)
      g. Calibration training (2.79, 3.98)
      h. Aggregation of expert opinion (3.34, 4.24)
      i. Immersive observation of AI labs (3.84, 3.03)
      j. **<u>Identifying clear and effective forecasting targets (4.29, 3.67)</u>**

---

[37] Moreover, it was used to represent a more verbose description from the first round of the Delphi and may have done a poor job at this.
[38] Excluding hybrid methods which only received 2 responses.
[39] Interested readers should begin by referring to Bendis et als' (2020) introduction to neural forecasting.

k. Conceptual progress acceleration survey (i.e. using pairwise comparisons) (3.78, 3.7)

In general, experts' interest in judgmental forecasting techniques seems only fair to moderate: only six of the fourteen techniques scored above average on importance while about half scored above average for feasibility. While expert elicitation is broadly scored below average for importance, the Delphi and expert adjustment scored modestly above average yet both were scored to be very feasible. Blue-team/red-team exercises are also above average for both importance and feasibility, and, because it is an underexplored technique for purposes related to AI forecasting, should receive relatively high priority among the methods listed here. Immersive observation of AI labs was scored as important, but would be challenging in practice and scored poorly for feasibility. Likewise, conceptual progress acceleration surveys scored above average for importance but less than average for feasibility. However, this technique only received five responses and may not have been well understood by participants[40].

Finally, there was only one method scoring among the top third on importance, which scored slightly below average for feasibility. This method - identifying clear and effective forecasting targets - is less a method unto itself and more a subtask of elicitation because elicitations are of no value if the targets are not of high quality. The importance of effective forecasting targets was also highlighted in the questions, so it is not particularly surprising to see it appear again. While Dafoe's (2018) desiderata for forecasting targets offers an excellent start, this is a very difficult but feasible task. Consequently, we feel it should be a priority for researchers interested in judgmental forecasting techniques.

Notably, only one method in this class scored in the top third on importance. Compared to five methods from statistical models scoring in the top third, judgmental techniques as a whole seem to be considered less important by experts. However, there is some reason to doubt these numbers' validity concerning the value of expert judgement in the context of forecasting AI progress. For example, Philip Tetlock, one of the world's leading forecasting experts, has suggested that there is value in expert opinion for the purpose of AI forecasting (Tetlock and Labenz 2020). Yet, this is just one expert's view and should not be given more weight simply because Tetlock, an expert on judgmental forecasting techniques, has published two bestselling and award winning books on the topic. This seems like sound logic, however, these books denounce expert judgement and scenario analysis techniques (Tetlock 2006; Tetlock and Gardner 2016). Yet, despite his very public and extensive criticism of these techniques, he now supports their use for AI forecasting purposes. Ironically, Tetlock has demonstrated in his books that beliefs which are updated in light of new information are more accurate (Tetlock and Gardner 2016), suggesting that his updating of his strong beliefs against the use of these techniques for forecasting should be given more weight. Moreover, Gruetzemacher (2019a) has suggested that a holistic approach to forecasting may be more appropriate in the context of forecasting AI progress, and Tetlock's proposed full-inference-cycle tournaments (Tetlock 2017) are one of two examples in the literature considered as such.

## D3 Hybrid & Other

Hybrid methods were not a major focus from participants in response to the 1st round questionnaire, and only five respondents scored them in the second round. However, they were scored to be the most important technique by over half a standard deviation. Gruetzemacher's (2019) holistic framework also suggests that holistic approaches involving a variety of methods may be better suited for forecasting AI progress, thus, they are discussed here with some concrete suggestions for future work. There were also a number of methods that were mentioned in the 1st round that didn't fit appropriately into the two primary classes. These are also discussed here, including some concrete suggestions. Although this may seem to be the neglected category, there are many very valuable concrete suggestions included in this section. Moreover, two of these topics were among the highest scoring methods, thus, this section should not be overlooked. The four included methods are listed in the brief outline below.

---

[40] Interested readers can find a proposal for this method at: https://tinyurl.com/pairwise-comparisons.

    **C. Hybrid methods (i.e. statistical and judgmental) (4.52, 3.78)**
    D. Other:
        a. Probabilistic reasoning (e.g. the Doomsday argument) (3.25, 3.0)
        b. **In-depth analysis of specific questions (4.19, 4.17)**
        c. **Literature review (3.88, *4.43*)**

As noted, hybrid methods scored significantly higher for importance than any of the other methods while also scoring above average on feasibility. With the exception of probabilistic reasoning, which scored poorly on both measures, the remaining two methods in the other class scored very well for both importance and feasibility. In-depth analysis of specific questions scored the fourth highest for importance while also scoring above 4 for feasibility, suggesting that it be a priority. Literature review did not score as highly on importance, but was scored to be the most feasible method. Consequently, we conclude that it should also be among the highest priority research methods; we consider it a low hanging fruit, so, while there remain fruit easy to pick, we suggest that those interested in furthering the study of AI forecasting consider working prioritizing these because they can benefit the community more broadly.